\documentclass[fleqn,twoside]{article}
\usepackage{amsmath}
\usepackage{espcrc2}
\usepackage{bm}
\usepackage{color}
\usepackage{graphicx}
\usepackage{here}
\def\Journal#1#2#3#4{{#1} {\bf #2}, #3 (#4)}

\def\EPJA{{\em Eur. Phys. J.} A}

\def\NPA{{\em Nucl. Phys.} A}
\def\NPB{{\em Nucl. Phys.} B}
\def\PLB{{\em Phys. Lett.}  B}
\def\PR{\em Phys. Rept.}
\def\PRL{\em Phys. Rev. Lett.}
\def\PRC{{\em Phys. Rev.} C}
\def\PRD{{\em Phys. Rev.} D}

\def\JHEP{\em JHEP}
\def\hepex#1{arXiv:\:hep-ex/{#1}}
\def\hepph#1{arXiv:\:hep-ph/{#1}}
\newcommand{\ud}     {\mathrm{d}}

\newcommand{\Mev}    {\:\mathrm{MeV}}
\newcommand{\gevsq}  {\:\mathrm{GeV}^2}

\newcommand{\ceps}{\varepsilon}

\newcommand{\average}[1]{\left\langle{#1}\right\rangle}
\newcommand{\eq}[1]{Eq.(\ref{#1})}
\newcommand{\Eqs}[2]{Eqs.(\ref{#1}) and (\ref{#2})}
\newcommand{\eqs}[1]{Eqs.(\ref{#1})}
\newcommand{\sqw}{\sin^2{\theta_W}}
\newcommand{\sqqw}{\sin^4{\theta_W}}

\newcommand{\as}{\alpha_{\scriptscriptstyle S}}

\newcommand{\PMF}{\mathcal{P}_{\mathrm{MF}}}
\newcommand{\Pcor}{\mathcal{P}_{\mathrm{cor}}}

\title{%
Relations Between NC and CC Neutrino Structure Functions
for Nucleons and Nuclei 
} 
\author{S. A. Kulagin
\address{
Institute for Nuclear Research, 117312 Moscow, Russia
}}
\begin{document}

\begin{abstract}
\noindent
The relations between neutrino NC and CC structure functions and 
cross sections, which are driven by isospin symmetry, are discussed for 
nucleons and nuclei. 
%\vspace{1pc} 
\end{abstract}

\maketitle

%%%%%%%%%%%%%%%%%%%%%%%%%%%%%%%%%%%%%%%%%%%%%%%%%%
%    BEGINNING OF TEXT                           %
%%%%%%%%%%%%%%%%%%%%%%%%%%%%%%%%%%%%%%%%%%%%%%%%%%
\section{%
Introduction
}

The studies of neutrino interactions with nuclei has been an actively 
developing field both experimentally and theoreticaly \cite{nuint,nuF}. 
Nuclear physics studies with (anti)neutrino beams are relevant in two 
different contexts. First, it must be noted that neutrino collision 
experiments usually use heavy-nucleus targets. For this reason a good 
understanding of ``standard'' nuclear effects is necessary for 
interpretation of high precision experiments with neutrino beams. On the 
other hand, neutrino experiments can provide better understanding of 
physics of nuclei that is important and interesting per se \cite{minerva}.

The interaction of (anti)neutrino with matter is mediated by charged $W^+$ 
or $W^-$ boson (charged current, CC), or by neutral $Z$ boson (neutral 
current, NC). For this reason the studies of neutrino interactions provide 
information which is not accessable with charged-lepton probes.
For example, the neutrino NC interaction strength is determined by the 
weak mixing angle $\theta_W$ and the studies of relative rates of NC 
and CC neutrino reactions provide a tool for the measurement of $\sqw$ 
\cite{nutev-prl}. 

In this paper we analyse relations between CC and NC neutrino structure 
functions for nucleons and nuclei. 
In Sect.~\ref{nc_cc:sec} we discuss the derivation of NC--CC relations for 
structure functions at high momentum transfer $Q$ for a generic target. 
First we discuss the structure functions in the QCD leading order in 
strong coupling constant (LO) in terms of parton distributions (PDFs) 
focusing on the analysis of contributions from different isospin states. 
We point out simple relations (see \eqs{nc-cc:SF}) emerging in this 
approximation. Then we discuss perturbative QCD corrections to these 
relations. 
In Sect.~\ref{nuke} we discuss nuclear corrections to parton distributions 
with different isospin.
The results are applied to compute the non-isoscalarity 
correction to the ratio $F_3^Z/F_3^W$ for iron.

\section{%
%RELATIONSHIPS BETWEEN CC AND NC NEUTRINO SCATTERING
NC/CC ratios for neutrino scattering
}
\label{nc_cc:sec}

For an isoscalar target (e.g. the isoscalar combination of proton 
and neutron, or for deuterium) a relation between 
neutrino--antineutrino asymmetries in the NC and CC deep-inelastic (DIS) 
cross sections was derived long ago by Paschos and Wolfenstein \cite{PW73}
\begin{equation}
\label{pw}
R^- =
\frac{
\sigma^\nu_{\mathrm{NC}}-\sigma^{\bar\nu}_{\mathrm{NC}}
}
{
\sigma^\nu_{\mathrm{CC}}-\sigma^{\bar\nu}_{\mathrm{CC}}
}
=\frac12 -\sqw,
\end{equation}
where $\theta_W$ is the Weinberg mixing angle.
The corresponding relation was also derived for the $C$-even combinations 
of cross sections \cite{LS83} (Llewellyn-Smith relationship)
\begin{equation}
\label{ls}
R^+ =
\frac{
\sigma^\nu_{\mathrm{NC}}+\sigma^{\bar\nu}_{\mathrm{NC}}
}
{
\sigma^\nu_{\mathrm{CC}}+\sigma^{\bar\nu}_{\mathrm{CC}}
}
=\frac12 -\sqw + \frac{10}{9}\sqqw.
\end{equation}

It should be remarked that in the derivation of the Paschos--Wolfenstein 
(PW) and the Llewellyn-Smith (LS) relationships the 
contributions from the $s$ and $c$ quarks were neglected. 
Furthermore, the derivation of the LS relationship holds for high momentum 
transfer $Q^2$ and this relationship should be corrected for perturbative 
and non-perturbative QCD effects even in an ideal world 
with only $u$ and $d$ quarks.
The PW relationship is, however, more general. If only the contributions 
due to light quarks are taken into account, the PW relationship is a 
direct result of the isospin symmetry. This ensures that various strong 
interaction effects, including nuclear effects, cancel out in $R^-$ for an 
isoscalar target thus making \eq{pw} a good tool for the measurement of 
the mixing angle in neutrino scattering.

If $s$ and $c$ quarks are taken into account the $C$-even ratio $R^+$ 
involves contributions due to $s+\bar s$ and $c+\bar c$ distributions, 
while the $C$-odd $R^-$ is corrected by $s-\bar s$ and $c-\bar c$ 
asymmetries in the target (for a discussion of possible asymmetry in the 
strange sea and the magnitude of this correction to $R^-$ see 
\cite{DFGRS02,mm03,Kretzer:2003wy,Portheault:2004xy}). 
It should be remarked that relations 
(\ref{pw}) and (\ref{ls}) are also violated by isospin-violating effects 
in PDFs (for a recent discussion of this effect see 
\cite{Londergan:2003pq}).

The targets used in neutrino experiments are usually heavy nuclei, such as 
iron in NuTeV experiment~\cite{nutev-prl}. Heavy nuclei typically have an 
excess of neutrons over protons and, therefore, are not isoscalar targets. 
For a non-isoscalar target relations (\ref{pw}) and (\ref{ls}) are 
violated by contributions due to isovector component of PDFs.

In this paper we consider the ratios $R^\pm$ for inclusive differential 
cross sections for a generic target as a function of Bjorken $x$ and the 
energy transfere in units of beam energy $E$, $y=q_0/E$. In terms of the 
structure functions $F_2$, $F_3$, and $F_L$ we have 
\begin{align}
\label{Rminus}
R^- &= \frac{xF^Z_3 Y_- /2} {xF^W_3 Y_-  + \Delta F_2^W Y_+ 
- \frac12 \Delta F_L^W y^2},\\
R^+ &= \frac{(F_2^Z Y_+ + \frac12 F_L^Z y^2)/2} {F_2^W Y_+ + \frac12 
F_L^W y^2 + \Delta xF^W_3 Y_-}, 
\label{Rplus}
\end{align}
where $Y_\pm=\frac12[1\pm (1-y)^2]$, the superscript $Z$ and $W$ label the 
NC and CC neutrino structure functions, respectively, 
$F^W_2=\frac12(F^\nu_2+F^{\bar\nu}_2)$ and 
$\Delta F_2^W=\frac12(F^\nu_2-F^{\bar\nu}_2)$ 
and similar definitions for $F_3$ and the longitudinal structure function 
$F_L$ (in \Eqs{Rminus}{Rplus} we neglect the factor 
$(1+Q^2/M_W^2)^2/(1+Q^2/M_Z^2)^2$ arising due to the ratio of propagators 
of $W$ and $Z$ whose effect is small for feasible $Q^2$).

\subsection{Relations between CC and NC structure functions}
\label{nc_cc:subsec}

We now address relations between the NC and CC structure functions. We 
assume that $Q^2$ is high enough to apply the leading twist (LT) QCD 
approximation. In this approximation the NC and CC structure functions are 
given in terms of PDFs. In order to facilitate discussion of isospin 
effects, we consider the isoscalar, $q_0(x)=u(x)+d(x)$, and the isovector, 
$q_1(x)=u(x)-d(x)$, quark distributions (for simplicity of notations, we 
suppress the explicit notation for the $Q^2$ dependence of parton 
distributions). We also introduce quark distributions with definite 
isospin $I=0,1$ and $C$ parity 
\begin{equation}
F^{(I,\pm)}(x) = xq_I(x) \pm x\bar q_I(x) , 
\label{F:IC}
\end{equation}
where $\bar q$ is antiquark distribution. We first consider the CC 
structure functions. In terms of the functions $F^{(I,C)}$ the structure 
functions can be written as follows (in QCD leading order) 
\begin{align}
\label{sf:CC}
\begin{split}
F_2^W &= F^{(0,+)}+ F^{(s,+)}+F^{(c,+)},\\
xF_3^W &= F^{(0,-)}+F^{(s,-)}+F^{(c,-)},\\
\Delta F_2^W &= -F^{(1,-)} + F^{(s,-)}-F^{(c,-)},\\
\Delta xF_3^W &= -F^{(1,+)} + F^{(s,+)}-F^{(c,+)}. 
\end{split}
\end{align}
where for $C$-even and $C$-odd distributions of strange ($I=s$) and 
charmed ($I=c$) quarks we use notations similar to \eq{F:IC}.

For the neutrino NC scattering the LO structure functions can be written 
as
\begin{align}
\label{sf:NC}
\begin{split}
F_2^Z ={ }& C_2^0 F^{(0,+)}+C_2^1 F^{(1,+)}+\\
       & C_2^s F^{(s,+)}+C_2^c F^{(c,+)},
\\
xF_3^Z ={ }& C_3^0 F^{(0,-)}+C_3^1 F^{(1,-)}+\\
        & C_3^s F^{(s,-)}+C_3^c F^{(c,-)}.
\end{split}
\end{align}
The coeficients $C_2^I$ and $C_3^I$ in \eqs{sf:NC} are
\begin{align} 
\label{C2}
\begin{split}
C_2^0 &=  1-2\sqw + \tfrac{20}{9} \sqqw ,
\\
C_2^1 &=  -\tfrac{2}{3}\sqw (1-2\sqw),
\\
C_2^s &=  1- \tfrac{4}{3} \sqw + \tfrac{8}{9} \sqqw,
\\
C_2^c &=  1- \tfrac{8}{3} \sqw + \tfrac{32}{9} \sqqw,
\end{split}
\end{align}
\begin{align}
\label{C3}
\begin{split}
C_3^0 &=  1-2 \sqw,
\\
C_3^1 &=  - \tfrac{2}{3} \sqw,
\\
C_3^s &=  1- \tfrac{4}{3} \sqw,
\\
C_3^c &=  1- \tfrac{8}{3} \sqw.
\end{split}
\end{align}
Using equations (\ref{sf:CC}) to (\ref{C3}) we arrive at the 
following relations between CC and NC structure functions
\begin{subequations}
\label{nc-cc:SF}
\begin{align}
\label{nc-cc:2}
F_2^Z &= C_2^0 F_2^W - C_2^1 \Delta xF_3^W,\\
xF_3^Z &= C_3^0 xF_3^W - C_3^1 \Delta F_2^W.
\label{nc-cc:3}
\end{align}
\end{subequations}

These relations have been derived in the LO 
approximation. 
It is important to study QCD 
corrections to these relations.
This is most conveniently done in the DIS scheme 
\cite{aem78}, in which only $F_L$ and $F_3$ structure functions change.
In this scheme perturbative corrections to \eqs{nc-cc:SF} are 
determined by the corresponding correction to the structure function $F_3$. 
Let us denote for each of the structure functions $i=2,3,L$ $F_i=F_i^{\rm 
LO} + \delta F_i$, where $F_i^{\rm LO}$ are the LO structure functions 
by \Eqs{sf:CC}{sf:NC} and $\delta F_i$ are perturbative series in $\as$
which can be written as convolutions of coefficient functions with LO 
structure functions. For the structure function $F_3$
\begin{equation}
\label{dF3}
\delta F_3 = \int_x^1\frac{\ud z}{z}K_3(z,\as(Q^2))
F_3^{\rm LO}(\frac{x}{z},Q^2),
\end{equation}
where $K_3$ is the corresponding coefficient function.
In the DIS scheme $\delta F_2^{Z,W}=0$ to all order in perturbation theory. 
Therefore, the generalization of \eq{nc-cc:2} can be written as
\begin{subequations}
\label{nc-cc:SF:cor}
\begin{equation}
\label{nc-cc:2:cor}
F_2^Z - C_2^0 F_2^W 
= -C_2^1 \Delta x F_3^{W({\rm LO})}(x,Q^2) ,
\end{equation}
where from the left we have the full ($\as$-cor\-rect\-ed) structure 
functions and from the right the expression by \eq{sf:CC} should be used, 
that is indicated by the superscript LO.

The generalization of \eq{nc-cc:3} is somewhat more complex. We use 
\Eqs{dF3}{nc-cc:3} for the LO structure functions in order to find
\begin{align}
\begin{split}
xF_3^Z - C_3^0 xF_3^W =
- C_3^1 \biggl( \Delta F_2^{W({\rm LO})} + {} &\\
  \int_x^1\ud z K_3(z,\as)
        \Delta F_2^{W({\rm LO})}\big(\frac{x}{z},Q^2\big)\biggr) &,
\end{split}
\label{nc-cc:3:cor}
\end{align}
\end{subequations}
where, similar to \eq{nc-cc:2:cor}, from the left we have the full 
structure functions and from the right the parton-model expression should 
be used.

Equations (\ref{nc-cc:2}) and (\ref{nc-cc:3}) [as well as  
$\as$-cor\-rect\-ed \Eqs{nc-cc:2:cor}{nc-cc:3:cor}] hold for a generic 
target, not necessarily isoscalar. We note that strong interaction 
corrections come through the isosvector PDF and/or $s$ or $c$ quark 
distribution. For an isoscalar target the isovector distributions vanish 
$F^{(1,\pm)}=0$, provided that the isospin symmetry is exact. If we 
further neglect the contributions from $s$ and $c$ quarks, then the 
asymmetries $\Delta F_{2,3}$ vanish and \eqs{nc-cc:SF:cor} reduce to the 
LS and PW relationships for the structure functions
\begin{subequations}
\label{ls-pw:SF}
\begin{align}
\label{ls:SF}
F_2^Z/F_2^W  &= C_2^0 ,\\
F_3^Z/F_3^W  &= C_3^0 .
\label{pw:SF}
\end{align}
\end{subequations}
Relationships (\ref{pw}) and (\ref{ls}) follow if we additionaly neglect 
contributions from the longitudinal structure function.
It must be remarked, however, that the accuracy of the
relations in the $C$-odd channel is higher, because they involve 
asymmetries $s-\bar s$ and $c-\bar c$, while the $C$-even relations are 
strongly corrected by strange quark effect at small $x$. 
Note also that if 
$s=\bar s$ and $c=\bar c$ (which is widely employed assumption in 
PDF analyses) then $\Delta F_{2,L}^W=0$ and \Eqs{pw:SF}{pw} hold to all 
orders in $\as$ for the isoscalar target, as it follows from 
\eq{nc-cc:3:cor}.
We also note that $\mathcal{O}(\as)$ corrections for the NC/CC ratios 
of (anti)neutrino cross sections for the isoscalar target were recently 
discussed in Ref.\cite{Dobrescu:2003ta}, in which it was argued that the 
$\as$ correction is suppressed by a factor of $\sqqw$. For the full 
analysis one has to take into account electro-weak corrections as well 
\cite{Diener:2003ss,Arbuzov:2004zr}.

If the target is not isoscalar then isovector quark distributions 
$F^{(1,\pm)}$ may be finite, which is the case for complex nuclei. The 
isovector quark distributions 
violate the LS and PW relationships for structure functions and cross 
sections. In order to address this effect, in Sect.~\ref{nuke} we 
discuss nuclear effects for the isoscalar and isovector quark 
distributions.

\section{%
%NUCLEAR EFFECTS FOR QUARK DISTRIBUTIONS WITH ISOSPIN 0 AND 1
Nuclear PDFs with isospin 0 and 1
}
\label{nuke}

Complex nuclei, such as iron, have unequal number of neutrons ($N$) and 
protons ($Z$), and the isovector quark distribution is finite in such 
nuclei. In order to quantitatively understand this effect, we denote 
$q_{a/T}$ as the distribution of quarks of type $a$ in a target $T$.
If $x$ is large enough to neglect coherent nuclear shadowing effect,
the lepton scattering off a nucleus can be well approximated as 
incoherent scattering off bound protons and neutrons (for a recent review 
of nuclear deeply inelastic scattering see \cite{PW99}). A widely used 
approximation is to neglect final state interactions of produced hadronic 
states with recoiling nucleus. In this approximation the nuclear parton 
distribuitions (nPDF) can be written as 
\begin{equation}
q_{a/A} = \average{q_{a/p}}_p + \average{q_{a/n}}_n, 
\label{nuke:qA}
\end{equation}
where the  two terms in the right side are the quark distributions 
in bound protons and neutrons averaged with the proton and neutron 
nuclear spectral functions, respectively. Similar equations can also be 
written for antiquark distributions. The explicit expression for the 
averaging in \eq{nuke:qA} is (for derivation and more details see 
\cite{ku89,kpw94,kp04}) 
\begin{align}
\label{nuke:av}
\begin{split}
x\langle q_{a/p}\rangle_p =& 
\int \ud\ceps\ud^3\bm{k}\, 
\mathcal{P}_p(\ceps,\bm{k}) \left(1+{k_z}/{M}\right) 
\\
& \qquad\qquad
x'q_{a/p}(x',Q^2,k^2),
\end{split}
\\
x'=&\frac{Q^2}{2k\cdot q}=\frac{x}{1+(\ceps+k_z)/M}. 
\label{xprim}
\end{align}
The integration in \eq{nuke:av} is taken over the energy  and momentum of 
bound protons (we separate the nucleon mass $M$ from the nucleon 
energy $k_0=M+\ceps$). The quantity $\mathcal{P}_p(\ceps,\bm{k})$ is the 
nuclear spectral function which describes the distribution of bound 
protons over the energy and momentum in nuclear ground state. The spectral 
functions $\mathcal{P}_{p}$ and $\mathcal{P}_{n}$ are normalized to the 
proton and neutron number, respectively. In \eq{nuke:av}, the $z$-axis is 
chosen in the direction opposite to the momentum transfer 
$q=(q_0,0_\perp,-|\bm{q}|)$, and $x'$ is the Bjorken variable of the bound 
proton with four-momentum $k$. Since bound nucleons are off-mass-shell 
particles their quark distributions generally depend on nucleon virtuality 
$k^2$ as additional variable 
(off-shell effect in structure functions is discussed in terms of 
different approaches in
~\cite{Gross:1991pi,Melnitchouk:1993nk,kpw94,ku98,akl04,kp04}).
Equation similar to \eq{nuke:av} also holds for neutrons 
with the obvious replacement of the spectral function and quark 
distributions.

Equations (\ref{nuke:qA}) and (\ref{nuke:av}) account for nuclear binding 
and Fermi motion effects in nPDF (for this effect we will use the 
abbreviation FMB). For the isoscalar and isovector nuclear parton 
distributions we obtain from Eq.(\ref{nuke:qA}) 
\begin{subequations}
\label{qA:01}
\begin{align}
\label{nuke:q0}
q_{0/A} &= \average{q_{0/p}}_0,\\
q_{1/A} &= \average{q_{1/p}}_1, 
\label{nuke:q1}
\end{align}
\end{subequations}
where the averaging is respectively performed with isoscalar and isovector 
spectral functions, $\mathcal{P}_0=\mathcal{P}_{p}+\mathcal{P}_{n}$ and 
$\mathcal{P}_1=\mathcal{P}_{p}-\mathcal{P}_{n}$.

The isoscalar and isovector spectral functions $\mathcal{P}_0$ and
$\mathcal{P}_1$ are very different in complex nuclei.
In an isoscalar nucleus with equal number of protons and
neutrons one generally assumes vanishing $\mathcal{P}_1$
and nuclear effects are dominated by the isoscalar spectral function.
In a nuclear mean-field model, in which a nucleus is viewed as Fermi gas
of nucleons bound to self-consistent mean field, the spectral function can
be calculated as
\begin{equation}
\label{spfn:MF}
\PMF(\ceps,\bm{p})=\sum_{\lambda<\lambda_F}
n_\lambda \left|\phi_\lambda(\bm{p})\right|^2
\delta(\ceps - \ceps_\lambda),
\end{equation}
where $\phi_\lambda(\bm{p})$ is the wave function of the
single-particle level $\lambda$ in nuclear mean field and $n_\lambda$ is
the number of nucleons on this level. The sum in \eq{spfn:MF} runs
over occupied single-particle levels with energies below the Fermi level
$\lambda_F$.
Equation (\ref{spfn:MF}) gives a good approximation to nuclear spectral
function in the vicinity of the Fermi level, where the excitation energies
of the residual nucleus are small.
As the separation energy $|\ceps|$ becomes higher, the 
mean-field approximation becomes less accurate. High-energy 
and high-momentum component of nuclear spectrum can not be described by 
the mean-field model and driven by correlation effects in nuclear ground 
state as witnessed by numerous studies (see, e.g.,~\cite{ben94,CS96}). We 
denote this contribution to the spectral function as 
$\Pcor(\ceps,\bm{p})$.

For a generic nucleus the spectral function $\mathcal{P}_1$ determines 
the isovector nucleon distribution. We now argue that the strength of 
$\mathcal{P}_1$ for complex nuclei is peaked about the Fermi surface. It is 
reasonable to assume that $\Pcor$ is mainly isoscalar and neglect its 
contribution to $\mathcal{P}_1$. Then $\mathcal{P}_1$ is determined by the 
difference of the proton and neutron mean-field spectral functions. If we 
further neglect small differences between the energy levels of protons and 
neutrons then $\mathcal{P}_1$ will be determined by the difference in the 
level occupation numbers $n_\lambda$ for protons and neutrons. Because of 
Pauli principle, an additional particle can join a Fermi system only on an 
unoccupied level. In a complex nucleus all but the Fermi level are usually 
occupied (the Fermi level has a large degeneracy factor). Therefore, 
$\mathcal{P}_1$ is determined by the contribution from the Fermi level and 
we can write
\begin{equation}
\label{spfn_1}
\mathcal{P}_1=(Z-N)|\phi_{F}(\bm{p})|^2\delta(\ceps-\ceps_F),
\end{equation}
where $\ceps_F$ and $\phi_F$ are the energy and the wave function of
the Fermi level.

\begin{figure}[htb]
%\vspace{9pt}
\begin{center}
\includegraphics[width=0.5\textwidth]{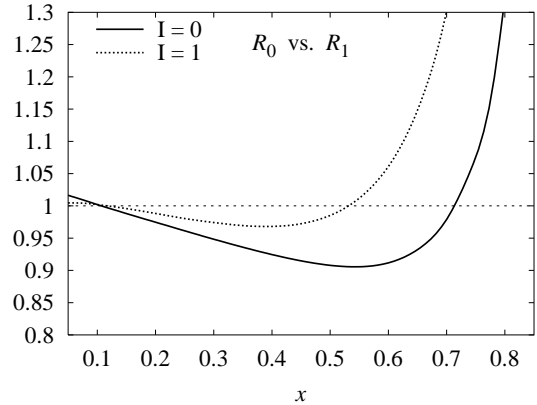}
\end{center}
%\vspace{-3em}
\caption{
Fermi motion and nuclear binding effects for isoscalar and 
isovector quark distributions in ${}^{56}$Fe${}_{26}$ calculated in 
impulse approximation for $Q^2=15\gevsq$. 
}
\label{fig:01}
\end{figure}

Figure \ref{fig:01} illustrates isospin dependence of nuclear effects for 
valence quark distributions. Shown are the 
ratios $R_0^A=F_A^{(0,-)}/(AF_p^{(0,-)})$ and 
$R_1^A=F_A^{(1,-)}/[(Z-N)F_p^{(1,-)}]$ calculated for the nucleus 
${}^{56}$Fe${}_{26}$. 
For the isoscalar nuclear spectral function we use the model spectral 
function of Ref.\cite{CS96} (see also \cite{KS00} for a different way 
of fixing parameters of the spectral function) which takes into account 
both the mean-field and correlated contributions. The isovector spectral 
function was calculated by \eq{spfn_1} using the Fermi gas model in which 
the wave function of the Fermi level $|\phi_F(p)|^2=\delta(p-p_F)/(4\pi 
p_F^2)$, where $p_F$ is the Fermi momentum. In numerical estimates we use 
$\ceps_F=-10\Mev$ and $p_F=260\Mev$ for the Fermi energy and momentum.

%%%
%%% FIG.1
%%%

As an application of the present analysis we calculate the isovector 
correction to relation (\ref{pw:SF}). 
Using Eq.(\ref{nc-cc:3:cor}) and neglecting $\as$ correction we obtain 
for the ratio of NC and CC nuclear structure functions
\begin{equation}
\frac{F_3^Z}{F_3^W} = 1-\sqw \left(2-\tfrac23 \delta N R^A_{1/0}(x)\right),
\label{pw:SF:cor}
\end{equation}
where $\delta N=(N-Z)/A$ is fractional neutron excess in a nucleus and
$R^A_{1/0}$ is the ratio of reduced 
nuclear isovector and isoscalar quark distributions, 
$[F_A^{(1,-)}/(Z-N)]/(F_A^{(0,-)}/A)$. The latter can also be written in 
terms of the corresponding ratio for the proton 
$R^p_{1/0}=F_p^{(1,-)}/F_p^{(0,-)}$ and the ratios $R_1$ and $R_0$ as 
$R^A_{1/0}=R^p_{1/0} R^A_1/R^A_0$.
Figure 2 shows the $x$ dependence of $R^A_{1/0}$ for the iron 
nucleus and the ratio $R^p_{1/0}$ for the proton. 

%%%
%%% FIG.2
%%%
\begin{figure}[htb]
%\vspace{-5em}
%\vspace{9pt}
\begin{center}
\includegraphics[width=0.5\textwidth]{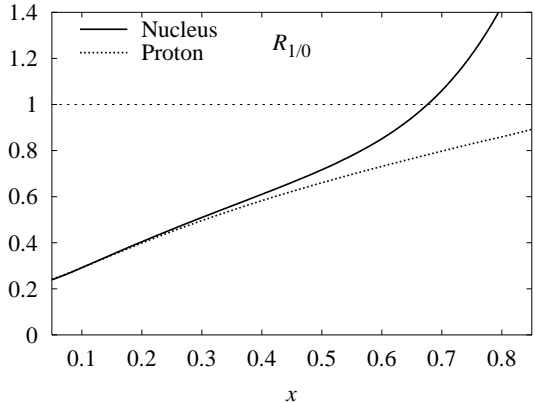}
\end{center}
\caption{
The ratio of isovector and isoscalar quark distributions
calculated for the proton and iron at $Q^2=15\gevsq$.
}
\label{fig:1_0}
\end{figure}

The parameter controlling the isovector correction in \eq{pw:SF:cor} is 
$\delta N$. This correction is additionally suppressed by the factor of 
$\sqw$. Given that $\delta N\approx 0.07$ for iron we observe that the 
isovector correction in \eq{pw:SF:cor} is not large. In particular, in the 
valence quark region of $x\sim 0.3$ the magnitude of the correction
in \eq{pw:SF:cor} is about 1\% of the PW value $1-2\sqw$. This correction 
rises with $x$, as is clear from Fig.~\ref{fig:1_0}. However, its 
magnitude is only about 2\% at $x=0.7$.

We note that the non-isoscalarity correction to $R^-$ for the total cross 
sections was recently discussed in \cite{ku03,ku04}. It was observed 
that the proper treatment of the FMB effect leads to about 6\% enhancement 
of the neutron excess correction for iron nucleus. We also remark that a 
subtle cancellation of perturbative QCD effects in the isovector 
correction for $R^-$ was found in \cite{ku04}. 

In summary, we discussed relations between NC and CC structure functions 
for a generic nuclear target. We argued that the relation 
$F_3^Z/F_3^W=1-2\sqw$ survives strong interaction corrections if the 
strange sea in the target is symmetric and the target is isoscalar.
The corresponding relation for $F_2$ is less accurate and affected
by strange quark effect at small $x$. 
We also discussed how the parton distributions with isospin 0 and 1 change 
in complex nuclei and applied the results to compute the non-isoscalarity 
correction to the ratio $F_3^Z/F_3^W$. 

\section*{%
Acknowledgments
}
This work was partially supported by the Russian Foundation for 
Basic Research projects no. 03-02-17177.
I am grateful to the organizers of the NuInt04 meeting 
for local support.


\begin{thebibliography}{99}
\bibitem{nuint}
Proceedings of Conferences of \emph{NuInt} series.
%
\bibitem{nuF}
M.~L.~Mangano, \emph{et al.}, \hepph{0105155}.
%
\bibitem{minerva}
Minerva Collaboration (D. Drakoulakos \emph{et al}.), \hepex{0405002}.
%
\bibitem{nutev-prl}
G.~P.~Zeller, \emph{et al.}, \Journal{\PRL}{88}{091802}{2002}.
%
\bibitem{PW73}
E.~A.~Paschos and L.~Wolfenstein, \Journal{\PRD}{7}{91}{1973}.
%
\bibitem{LS83}
C.~H.~Llewellyn-Smith, \Journal{\NPB}{228}{205}{1983}.
% 
\bibitem{DFGRS02}
S.~Davidson, S.~Forte, P.~Gambino, N.~Rius, and A.~Strumia,
\Journal{\JHEP}{02}{037}{2002}.
%
\bibitem{mm03}
K.~S.~McFarland and S.-O.~Moch, \hepph{0306052}.
%
%\cite{Kretzer:2003wy}
\bibitem{Kretzer:2003wy}
S.~Kretzer, F.~Olness, J.~Pumplin, D.~Stump, W.~K.~Tung and M.~H.~Reno,
%``The parton structure of the nucleon and precision determination of the
%Weinberg angle in neutrino scattering,''
%Phys.\ Rev.\ Lett.\  {\bf 93} (2004) 041802
\Journal{\PRL}{93}{041802}{2004}.
%[arXiv:hep-ph/0312322].
%%CITATION = HEP-PH 0312322;%%
%
%\cite{Portheault:2004xy}
\bibitem{Portheault:2004xy}
B.~Portheault,
%``Strange sea asymmetry from global QCD fits,''
arXiv:hep-ph/0406226.
%%CITATION = HEP-PH 0406226;%%
%
%\cite{Londergan:2003pq}
\bibitem{Londergan:2003pq}
J.~T.~Londergan and A.~W.~Thomas,
%``Charge symmetry violating contributions to neutrino reactions,''
\Journal{\PLB}{558}{132}{2003}.
%[arXiv:hep-ph/0301147].
%%CITATION = HEP-PH 0301147;%%
%
\bibitem{aem78}
G.~Altarelli, R.~K.~Ellis, and G.~Martinelli, 
\Journal{\NPB}{143}{521}{1978}; Erratum--ibid. B{\bf 146}, 544 (1978).
%
%\cite{Dobrescu:2003ta}
\bibitem{Dobrescu:2003ta}
B.~A.~Dobrescu and R.~K.~Ellis,
%``Analytic estimates of the QCD corrections to neutrino nucleus scattering,''
\Journal{\PRD}{69}{114014}{2004}.
%[arXiv:hep-ph/0310154].
%%CITATION = HEP-PH 0310154;%%
%
%\cite{Diener:2003ss}
\bibitem{Diener:2003ss}
K.~P.~O.~Diener, S.~Dittmaier and W.~Hollik,
%``Electroweak radiative corrections to deep-inelastic neutrino scattering:
%Implications for NuTeV?,''
\Journal{\PRD}{69}{073005}{2004}.
%[arXiv:hep-ph/0310364].
%%CITATION = HEP-PH 0310364;%%
%
%\cite{Arbuzov:2004zr}
\bibitem{Arbuzov:2004zr}
A.~B.~Arbuzov, D.~Y.~Bardin and L.~V.~Kalinovskaya,
%``Radiative corrections to neutrino deep inelastic scattering revisited,''
arXiv:hep-ph/0407203.
%%CITATION = HEP-PH 0407203;%%
%
\bibitem{PW99}
G.~Piller and W.~Weise,
%``Nuclear deep-inelastic lepton scattering and coherence phenomena,''
\Journal{\PR}{330}{1}{2000}.
%[arXiv:hep-ph/9908230].
%%CITATION = HEP-PH 9908230;%%
%
%\bibitem{akv85}
%S. V. Akulinichev, S. A. Kulagin, and G. M. Vagradov,
%\Journal{\PLB}{158}{485}{1985}.
%
\bibitem{ku89}
S.~A.~Kulagin, \Journal{\NPA}{500}{653}{1989}.
%
%\cite{Gross:1991pi}
\bibitem{Gross:1991pi}
F.~Gross and S.~Liuti,
%``Role of nuclear binding in the EMC effect,''
%Phys.\ Rev.\ C {\bf 45} (1992) 1374.
\Journal{\PRC}{45}{1374}{1992}.
%%CITATION = PHRVA,C45,1374;%%
%
%\cite{Melnitchouk:1993nk}
\bibitem{Melnitchouk:1993nk}
W.~Melnitchouk, A.~W.~Schreiber and A.~W.~Thomas,
%``Deep inelastic scattering from off-shell nucleons,''
%Phys.\ Rev.\ D {\bf 49} (1994) 1183
\Journal{\PRD}{49}{1183}{1994}.
%[arXiv:nucl-th/9311008].
%%CITATION = NUCL-TH 9311008;%%
% 
\bibitem{kpw94}
S.~A.~Kulagin, G.~Piller, and W.~Weise, \Journal{\PRC}{50}{1154}{1994}.
%
\bibitem{ku98}
S.~A.~Kulagin,  \Journal{\NPA}{640}{435}{1998}.
% 
\bibitem{akl04}
S.~I.~Alekhin, S.~A.~Kulagin and S.~Liuti,
\Journal{\PRD}{69}{114009}{2004}.
%
\bibitem{kp04}
S.~A.~Kulagin and R.~Petti, paper in preparation.
%
\bibitem{ben94}
O.~Benhar, A.~Fabrocini, S.~Fantoni, and I.~Sick,
\Journal{\NPA}{579}{493}{1994}.
%
\bibitem{CS96}
C.~Ciofi degli Atti and S.~Simula, \Journal{\PRC}{53}{1689}{1996}.
%
\bibitem{KS00}
S.~A.~Kulagin and A.~V.~Sidorov, \Journal{\EPJA}{9}{261}{2000}.
%
\bibitem{ku03}
S.~A.~Kulagin, \Journal{\PRD}{67}{091301(R)}{2003}.
%
\bibitem{ku04}
S.~A.~Kulagin, \hepph{0406220}.
%
%\bibitem{ku98}
%S. A. Kulagin, \hepph{9812532}.


\end{thebibliography}
\end{document}